\documentclass[preprint,nofootinbib,titlepage,prd]{revtex4}
\usepackage{amsfonts}
\usepackage{amsmath}
\usepackage{amssymb}
\usepackage{graphicx}

\input{tcilatex}

\begin{document}

\preprint{}
\title{Fermionic and Bosonic Stabilizing Effects\\
for Type I and Type II Dimension Bubbles}
\author{J.R. Morris}
\email{jmorris@iun.edu}
\affiliation{\textit{Physics Dept., Indiana University Northwest,3400 Broadway, Gary,
Indiana 46408 USA\bigskip }}

\begin{abstract}
We consider two types of \textquotedblleft dimension
bubbles\textquotedblright, which are viewed as 4d nontopological solitons
that emerge from a 5d theory with a compact extra dimension. The size of the
extra dimension varies rapidly within the domain wall of the soliton. We
consider the cases of type I (II) bubbles where the size of the extra
dimension inside the bubble is much larger (smaller) than outside. Type I
bubbles with thin domain walls can be stabilized by the entrapment of
various particle modes whose masses become much smaller inside than outside
the bubble. This is demonstrated here for the cases of scalar bosons,
fermions, and massive vector bosons, including both Kaluza-Klein zero modes
and Kaluza-Klein excitation modes. Type II bubbles expel massive particle
modes but both types can be stabilized by photons. Plasma filled bubbles
containing a variety of massless or nearly massless radiation modes may
exist as long-lived metastable states. Furthermore, in contrast to the case
with a \textquotedblleft gravitational bag\textquotedblright, the metric for
a fluid-filled dimension bubble does not exhibit a naked singularity at the
bubble's center.

PACS: 11.27.+d, 04.50.+h, 98.80.Cq
\end{abstract}

\maketitle

\section{Introduction}

An inhomogeneous compactification of a higher dimensional spacetime to four
dimensions may result in the formation of ``dimension bubbles'', where the
sizes of the extra dimensions inside such a bubble are much different than
those outside the bubble \cite{BG,JM,GM}. A dimension bubble, from a four
dimensional viewpoint, is a nontopological soliton consisting of a closed
domain wall that entraps particles and/or radiation, which help to stabilize
the bubble against collapse. Since the existence of dimension bubbles
depends only upon a dramatic change in the sizes of extra dimensions inside
and outside the bubble, the detections of such objects could provide
evidence for the existence of extra dimensions, regardless of how large or
small their ambient sizes. For simplicity and specificity, we will, as in
refs.\cite{JM,GM}, consider the case where there is one toroidally
compactified extra space dimension, so that the 5d spacetime has the
topology of $M_{4}\times S^{1}$.

Attention is focused here on several new features of dimension bubbles.
First, it is pointed out that two different types of dimension bubbles are
possible, and the size of the extra dimension, while differing dramatically
in the interior and exterior regions of the bubble, may remain microscopic
in both regions. Let us simply label these two bubble types as types I and
II. A type I (II) bubble is one for which the size of the extra dimension is
larger (smaller) inside the bubble than outside. A type I bubble can be
stabilized by massive particles that are trapped inside the bubble, where
the particle masses become much smaller than on the outside of the bubble.
This stabilization mechanism for type I bubbles was demonstrated for the
case of scalar bosons in ref.\cite{JM}. Here we extend these results to
include scalar bosons, fermions, and massive vector bosons--both
Kaluza-Klein (KK) zero modes and KK excitation modes. A type II bubble
expels such massive modes from its interior. The dependence of the particle
masses upon the extra dimensional scale factor $|\tilde{g}_{55}|^{1/2}=B(x)$
for the 4d Einstein frame is obtained in each case, for both the KK zero
modes and the KK excitations. Both type I and II bubbles can be stabilized
by photons, however, due to the high reflectivity of the bubble wall\cite{GM}%
.

Finally, we investigate the behavior of the 4d metric near the center of a
dimension bubble and compare it with that for a ``gravitational bag''\cite%
{DG,Guen}, which can be thought of as a static idealization of an empty type
I dimension bubble. (The gravitational bag solution solution is static, but
the scalar field of the bag exhibits a singular behavior at the bag's
center.) More specifically, we view a fluid-filled dimension bubble as a 
\textit{cosmic balloon}\cite{BH,wang} and investigate the behavior of the
interior metric near the bubble's center. Although the interior metric of a
``gravitational bag'' has a naked singularity near the geometric center\cite%
{DG,Guen}, it is seen that a fluid stabilized dimension bubble has a well
behaved, finite metric at the bubble's center. Therefore the naked
singularity of a gravitational bag is avoided in a fluid-filled dimension
bubble.

A brief summary of the dimension bubble model is presented in Sec.II and
conditions on the 4d effective potential are presented for the formation of
either type I or type II bubbles. We can consider as an example the case
where the extra dimension is, say, TeV sized (\textit{i.e.}, $l_{TeV}\sim $%
TeV$^{-1}$) in one region of space and Planck sized ($l_{P}\sim M_{P}^{-1}$)
in another region of space, so that the size of the extra dimension may
change by roughly 16 orders of magnitude between these two regions while
remaining microscopic in both regions. The mass dependence upon the extra
dimensional scale factor $|\tilde{g}_{55}|^{1/2}=B$ for KK zero modes is
exhibited in Sec. III for scalar bosons, fermions, and massive vector
bosons. The resulting expressions make clear how these particles can help
stabilize a type I bubble by getting trapped inside, as with an
\textquotedblleft ordinary\textquotedblright\ nontopological soliton of the
type previously studied by Frieman, Gleiser, Gelmini, and Kolb\cite{NTS},
and how they must be expelled from a type II bubble. These results are then
extended in Sec. IV to include Kaluza-Klein excitation modes. Radiation
filled metastable bubbles are then contemplated in Sec. V, where estimates
or rough bounds are obtained for bubble mass, radius, and lifetime. Finally,
gravitational aspects are addressed in Sec. VI, where it is pointed out that
by viewing a fluid filled dimension bubble as a \textit{cosmic balloon}\cite%
{BH} and using the results for a cosmic balloon metric\cite{wang}, the 4d
metric of a fluid-filled dimension bubble exhibits a nonsingular behavior at
the bubble's center. This suggests that the undesirable feature of a naked
singularity, which appears at the core of a \textquotedblleft gravitational
bag\textquotedblright \cite{DG,Guen}, does not appear at the center of a
fluid filled dimension bubble. Sec. VII consists of a brief summary.

\section{The Dimension Bubble Model}

\subsection{Metric Ansatz}

A five-dimensional (5d) spacetime is assumed to be endowed with a metric $%
\tilde{g}_{MN}$:

\begin{equation}
ds^{2}=\tilde{g}_{MN}dx^{M}dx^{N}=\tilde{g}_{\mu\nu}dx^{\mu}dx^{%
\nu}-B^{2}dy^{2}  \label{e1}
\end{equation}

\noindent where $x^{M}=(x^{\mu },y)$, with $M,N=0,\cdot \cdot \cdot ,3,5$, $%
\mu ,\nu =0,\cdot \cdot \cdot ,3$, and $B=\sqrt{-\tilde{g}_{55}}$ is the
dimensionless scale factor for the extra dimension. We assume an ansatz
where the metric $\tilde{g}_{MN}$ is independent of the extra dimension $y$,
i.e., $\tilde{g}_{MN}=\tilde{g}_{MN}(x^{\mu })$, $\partial _{5}\tilde{g}%
_{MN}=0$, and the metric factorizes with $\tilde{g}_{\mu 5}=0$. The extra
dimension, characterized by the coordinate $x^{5}=y$, with $0\leq y\leq 2\pi
R$, is taken to be toroidally compact, so that the 5d spacetime has a
topology of $M_{4}\times S^{1}$. We allow for the possibility that the scale
factor $B$ has a spatial dependence, i.e., $B=B(x^{\mu })$. In the
dimensionally reduced effective 4d theory the scale factor $B$ can be
associated with a scalar field $\varphi $ through the relation 
\begin{equation}
\varphi =\frac{1}{\kappa _{N}}\sqrt{\frac{3}{2}}\ln B,  \label{e2}
\end{equation}

\noindent where $\kappa_{N}$ is related to the 4d Planck mass $M_{P}$ by $%
\kappa_{N}=\sqrt{8\pi G}=\sqrt{8\pi}M_{P}^{-1}$, so that the scale factor
can be written as $B=e^{\sqrt{2/3}\,\kappa_{N}\varphi}$. We further consider
the situation wherein the scalar $\varphi$ is governed by a 4d effective
potential $U(\varphi)\geq0$, which arises from a Rubin-Roth potential for
bosonic and fermionic degrees of freedom\cite{RR}(see also ref.\cite{JM}),
along with a 5d cosmological constant $\Lambda$. When $U(\varphi)$ assumes a
``semi-vacuumless'' form characterized by the existence of a local minimum
at some finite value $\varphi=\varphi_{0}$, a local maximum at some finite
value $\varphi=\varphi_{\max}>\varphi_{0}$, and an asymptotic form $%
U(\varphi )\rightarrow U_{\infty}=const$ as $\varphi\rightarrow\infty$,
``dimension bubbles'' can arise\cite{JM,GM,DG} as solutions of the 4d theory
where the scalar $\varphi$ (and therefore the scale factor $B$) can vary
rapidly across a region of space. It is this rapid variation of $\varphi$
that is associated with the domain wall bounding the soliton. We will be
interested in the cases where $B$ differs dramatically in the interior and
exterior of the bubble. We note that even though the extra dimension may
remain microscopic in both regions, there can still be an enormous variation
in $B$ across the domain wall. If, for example, the extra dimension varies
across the bubble wall from a Planck size where $(BR)_{P}\sim$ $M_{P}^{-1}$,
characteristic of a ``small'' extra dimension, to a ``TeV size'' where $%
(BR)_{TeV}\sim TeV^{-1}$, which may be characteristic of a ``large'' extra
dimension, $B$ can change by 16 orders of magnitude.

\subsection{Dimensional Reduction of the 5d Action}

We take the 5-dimensional action to include the 5d Einstein action,
cosmological constant $\Lambda$, and a source Lagrangian $\mathcal{L}_{5}$: 
\begin{equation}
S_{5}=\dfrac1{2\kappa_{N(5)}^{2}} \int d^{5}x\sqrt{\tilde{g}_{5}}\left\{ 
\tilde{R}_{5}-2\Lambda+2\kappa_{N(5)}^{2}\mathcal{L}_{5}\right\}  \label{e3}
\end{equation}

\noindent where $\kappa_{N(5)}^{2}=8\pi G_{5}=(2\pi R)\kappa_{N}^{2}$, $%
\tilde{g}_{5}=|\det\tilde{g}_{MN}|$ and $\tilde{R}_{5}=\tilde{g}^{MN}\tilde{R%
}_{MN}$ denotes the 5-dimensional Ricci scalar built from $\tilde
{g}_{MN}$%
. A 4d Einstein Frame metric $g_{\mu\nu}$ can be defined in terms of the 4d
Jordan Frame metric $\tilde{g}_{\mu\nu}$ by $g_{\mu\nu}=B\tilde{g}%
_{\mu\nu}=e^{\sqrt{2/3}\kappa_{N}\varphi}\tilde{g}_{\mu\nu}$, and the line
element in (\ref{e1}) then takes the Kaluza-Klein form 
\begin{equation}
\begin{array}{ll}
ds^{2} & =B^{-1}g_{\mu\nu}dx^{\mu}dx^{\nu}-B^{2}dy^{2} \\ 
& =e^{-\sqrt{2/3}\,\kappa_{N}\varphi}g_{\mu\nu}dx^{\mu}dx^{\nu}-e^{2\sqrt {%
2/3}\,\kappa_{N}\varphi}dy^{2}%
\end{array}
\label{e4}
\end{equation}

Using (\ref{e3}) and (\ref{e4}), the 5d action is dimensionally reduced to
the effective 4d Einstein Frame action\cite{JM,GM} 
\begin{equation}
S=\int d^{4}x\sqrt{-g}\left\{ \frac1{2\kappa_{N}^{2}}R+\frac12(\nabla
\varphi)^{2}+e^{-\sqrt{2/3}\kappa_{N}\varphi}[\mathcal{L}-\frac1{\kappa
_{N}^{2}}\Lambda]\right\}  \label{e5}
\end{equation}

\noindent where $R=g^{\mu\nu}R_{\mu\nu}$ is the 4d Ricci scalar built from
the 4d Einstein Frame metric $g_{\mu\nu}$ and $g=\det g_{\mu\nu}$ and $%
\mathcal{L}=(2\pi R)\mathcal{L}_{5}$.

The 4d effective Lagrangian that is generated by $\mathcal{L}$ is $\mathcal{L%
}_{4}=B^{-1}\mathcal{L}$. The Lagrangian $\mathcal{L}_{4}$, the effective
potential $U(\varphi)$, the $\varphi$ kinetic term $\frac
12(\partial\varphi)^{2}$, and the gravitational term $\frac1{2%
\kappa_{N}^{2}}R$ produce a total 4d effective Lagrangian 
\begin{equation}
\begin{array}{ll}
\mathcal{L}_{eff} & =\frac1{2\kappa_{N}^{2}}R+\frac12(\partial\varphi
)^{2}-U(\varphi)+\mathcal{L}_{4}%
\end{array}
\label{e6}
\end{equation}

The \textquotedblleft semi-vacuumless\textquotedblright\ potential $%
U(\varphi )$ admits a domain wall solution separating a region where $B$
becomes very \textquotedblleft large\textquotedblright\ (where $\varphi $
assumes a value $\varphi _{1}>\varphi _{\max }$) from a region where $B$ is
relatively \textquotedblleft small\textquotedblright\ (at the local minimum
of $U$, where $\varphi =\varphi _{0}$). In general, $U(\varphi _{0})\neq
U(\varphi _{1})$ and the wall is unstable against bending toward the region
of higher energy density\cite{JM}, and we expect the formation of a network
of bubbles to result. A dimension bubble encloses a region of higher vacuum
energy density and is surrounded by a region of lower vacuum energy density.
For simplicity we consider a spherical thin walled bubble of radius $R_{B}$,
with wall thickness $\delta \ll R_{B}$, so that in a simplifying limit we
may take the inner radius $R_{-}$ and outer radius $R_{+}$ of the wall to
coincide, $R_{-}$, $R_{+}\rightarrow R_{B}$. It is within the wall that the
scale factor $B(x)$ varies rapidly, and we are interested in the case where $%
B$ takes on vastly different values in the interior and exterior regions.

\subsection{Type I and Type II Dimension Bubbles}

We can envision two distinct possibilities, corresponding to different sets
of model parameters, which can result in two different types of bubbles.
Either (i) $U(\varphi_{1})>U(\varphi_{0})$, giving rise to what we will
refer to as ``type I'' bubbles, or (ii) $U(\varphi_{1})<U(\varphi_{0})$
associated with ``type II'' bubbles. In the first case (type I) the bubble
interior contains a ``vacuum'' characterized by $\varphi\approx\varphi_{1}$,
a vacuum energy density $U(\varphi_{1})$, and a relatively large scale
factor $B\approx B_{1}=e^{\sqrt{2/3}\,\kappa_{N}\varphi_{1}}$ and in the
bubble's exterior region where $\varphi=\varphi_{0}$ there is a relatively
small scale factor $B_{0}=e^{\sqrt{2/3}\,\kappa_{N}\varphi_{0}}$. In the
second case (type II) we have the opposite situation. Therefore, a type I
bubble encloses a ``large'' extra dimension and is surrounded by a ``small''
extra dimension. A type II bubble encloses a ``small'' extra dimension and
is surrounded by a ``large'' extra dimension. Again, the extra dimension may
remain microscopic in all regions, but we entertain the possibility that its
size, characterized by $BR$, may be extremely different inside and outside
of a bubble. For example, we might consider the range $M_{P}^{-1}\lesssim
BR\lesssim TeV^{-1}$, in which case the values of $B$ inside and outside the
bubble would be related by $B_{in}/B_{out}\sim10^{\pm16}$.

\section{Effects of Kaluza-Klein Zero Modes}

In this section we consider contributions to the 4d effective Lagrangian $%
\mathcal{L}_{4}$ from Kaluza-Klein (KK) zero modes of scalar bosons,
fermions, and vector bosons which acquire mass through the Higgs mechanism.
Each type of zero mode particle field $\Phi$ is $x^{5}$--independent, i.e., $%
\Phi=\Phi(x^{\mu})$, $\partial_{5}\Phi=0$. We later consider Kaluza-Klein
excitations where the fields have a $y$--dependence from the cylinder
condition. The difference in size of the extra dimensional scale factor $B$
in the interior and exterior regions of the bubble results in a difference
in the effective particle mass in these regions. Specifically, the particle
mass becomes smaller in a region where $B$ is larger. This results in
particles getting trapped inside of type I bubbles and being expelled from
type II bubbles. Therefore the KK zero modes have a stabilizing influence on
type I bubbles, where the particle pressure can help to support the bubble
against collapse due to the wall tension. The dependence of the particle
mass $m$ upon $B$ is isolated for each type of particle.

\subsection{Scalar Bosons}

Consider a contribution to the Lagrangian $\mathcal{L}$ from a scalar boson $%
\phi$, 
\begin{equation}
\mathcal{L}_{S}=\tilde{\partial}^{M}\phi^{*}\tilde{\partial}_{M}\phi-\mu
_{0}^{2}\phi^{*}\phi=\tilde{g}^{\mu\nu}\partial_{\mu}\phi^{*}\partial_{\nu
}\phi-\mu_{0}^{2}|\phi|^{2}=B|\partial\phi|^{2}-\mu_{0}^{2}|\phi |^{2}
\label{e7}
\end{equation}

\noindent where $\partial_{5}\phi=0$ and $\tilde{g}^{\mu\nu}=Bg^{\mu\nu} $
and $|\partial\phi|^{2}=\partial^{\mu}\phi^{*}\partial_{\mu}\phi$. This
Lagrangian gives rise to the 4d effective Lagrangian 
\begin{equation}
\mathcal{L}_{4,S}=B^{-1}\mathcal{L}_{S}=|\partial\phi|^{2}-B^{-1}%
\mu_{0}^{2}|\phi|^{2}  \label{e8}
\end{equation}

\noindent The scalar boson mass in the effective 4d theory is therefore
identified as 
\begin{equation}
m_{S}=B^{-1/2}\mu_{0}  \label{e9}
\end{equation}

\noindent where $\mu_{0}$ is the mass parameter in the original 5d theory.
We see that since the value of $B$ inside of a type I bubble is assumed to
be much bigger than that outside the bubble, i.e., $B_{in}\gg B_{out}$, the
effective boson mass inside is relatively small, $m_{S,in}\ll m_{S,out}$.
The scalar boson is effectively trapped inside the type I bubble since there
is an enormous inward force $\vec{F}\approx-\nabla
m_{S}=-\mu_{0}\nabla(B^{-1/2})$ acting on the particle. The kinetic energies
of the light trapped particles exert an outward pressure on the bubble wall
to help stabilize it against collapse. However, for a type II bubble the
particle mass becomes much smaller outside the bubble, so that massive
particles that are initially present inside the bubble are expelled from it.

\subsection{Fermions}

Now consider a fermionic contribution to the Lagrangian $\mathcal{L}$ in the
form 
\begin{equation}
\mathcal{L}_{F}=\bar{\psi}^{\prime}\left( i\Gamma^{M}\partial_{M}-\mu
_{0}\right) \psi=\bar{\psi}^{\prime}\left(
i\Gamma^{\mu}\partial_{\mu}-\mu_{0}\right) \psi  \label{e10}
\end{equation}

\noindent where $\psi=\psi(x)$, $\partial_{5}\psi=0$, and $\bar{\psi}%
^{\prime }=\psi^{\dagger}\Gamma^{0}$. The $\Gamma^{M}$ matrices are taken to
be normalized according to 
\begin{equation}
\left\{ \Gamma^{M},\Gamma^{N}\right\} =-2\tilde{g}^{MN}  \label{e11}
\end{equation}

\noindent The 5d metric $\tilde{g}_{MN}$, written in terms of the 4d
Einstein frame metric $g_{\mu\nu}$ and the scale factor $B$, is 
\begin{equation}
\tilde{g}_{MN}=\left( 
\begin{array}{cc}
B^{-1}g_{\mu\nu} &  \\ 
& -B^{2}%
\end{array}
\right)  \label{e12}
\end{equation}

\noindent Eq. (\ref{e11}) then implies that $\left\{ \Gamma^{\mu},\Gamma
^{\nu}\right\} =-2Bg^{\mu\nu}.$

For the effective 4d Einstein Frame theory we define the new matrices $%
\gamma^{\mu}$ related to the original $\Gamma^{\mu}$ matrices by 
\begin{equation}
\Gamma^{\mu}=B^{1/2}\gamma^{\mu},\,\,\,\,\,\,\,\Gamma^{5}=B^{-1}\gamma ^{5}
\label{e13}
\end{equation}

\noindent with a normalization given by 
\begin{equation}
\left\{ \gamma^{\mu},\gamma^{\nu}\right\} =-2g^{\mu\nu},\,\,\,\,\,(\gamma
^{5})^{2}=1  \label{e14}
\end{equation}

\noindent Upon defining $\bar{\psi}=\psi^{\dagger}\gamma^{0}$ we have $\bar{%
\psi}^{\prime}=\psi^{\dagger}\Gamma^{0}=\psi^{\dagger}B^{1/2}\gamma
^{0}=B^{1/2}\bar{\psi}$, and the Lagrangian $\mathcal{L}_{F}$ can be
rewritten as 
\begin{equation}
\mathcal{L}_{F}=B\bar{\psi}\left( i\gamma^{\mu}\partial_{\mu}-B^{-1/2}\mu
_{0}\right) \psi  \label{e15}
\end{equation}

\noindent This Lagrangian gives rise to an effective 4d fermion Lagrangian 
\begin{equation}
\mathcal{L}_{4,F}=B^{-1}\mathcal{L}_{F}=\bar{\psi}\left( i\gamma^{\mu
}\partial_{\mu}-B^{-1/2}\mu_{0}\right) \psi  \label{e16}
\end{equation}

\noindent The fermion mass in the effective 4d theory is therefore
identified as 
\begin{equation}
m_{F}=B^{-1/2}\mu_{0}  \label{e17}
\end{equation}

\noindent which resembles the result obtained for scalar bosons.

\subsection{Massive Vector Bosons}

Let us consider the case of vector gauge bosons which acquire mass by the
Higgs mechanism, through the interaction with a scalar field $\chi$. There
is then a contribution to the Lagrangian given by 
\begin{equation}
\mathcal{L}_{G}=-\frac14\tilde{F}^{\prime MN}\tilde{F}_{MN}^{\prime}+(\tilde{%
D}^{M}\chi)^{*}(\tilde{D}_{M}\chi)|_{\chi=\eta}  \label{e18}
\end{equation}

\noindent where the tildes remind us that the metric $\tilde{g}_{MN}$ is
used to construct 5d scalars, so that, for instance, $(\tilde{D}^{M}\chi
)^{*}(\tilde{D}_{M}\chi)=\tilde{g}^{MN}(D_{N}\chi)^{*}(D_{M}\chi)$. The
field strength and gauge covariant derivative terms are given by 
\begin{equation}
F_{MN}^{\prime}=\partial_{M}A_{N}^{\prime}-\partial_{N}A_{M}^{\prime
},\,\,\,\,\,\,\,\,D_{M}\chi=(\nabla_{M}+ie_{0}A_{M}^{\prime})\chi
\label{e19}
\end{equation}

\noindent We choose $A_{5}^{\prime}=0$ and a vacuum state characterized by $%
\chi=\eta=const$. In the vacuum state we then have 
\begin{equation}
(\tilde{D}^{M}\chi)^{*}(\tilde{D}_{M}\chi)|_{\chi=\eta}=\tilde{g}%
^{MN}e_{0}^{2}\eta^{2}A_{M}^{\prime}A_{N}^{\prime}=\frac12B\mu_{0}^{2}A^{%
\prime\mu }A_{\mu}^{\prime}  \label{e20}
\end{equation}

\noindent where we have defined $\mu_{0}=\sqrt{2}e_{0}\eta$.

In order to obtain a canonical gauge field term in the 4d theory, we
introduce the gauge field $A_{\mu}=B^{1/2}A_{\mu}^{\prime}$. In terms of the
metric $g_{\mu\nu}$ and the gauge field $A_{\mu}$, we can rewrite $\mathcal{L%
}_{G}$ in the form 
\begin{equation}
\mathcal{L}_{G}=-\frac14BF^{\mu\nu}F_{\mu\nu}+\frac12\mu_{0}^{2}A^{\mu}A_{%
\mu }-\frac12B^{3/2}F^{\mu\nu}H_{\mu\nu}-\frac14B^{2}H^{\mu\nu}H_{\mu\nu }
\label{e21}
\end{equation}

\noindent where 
\begin{equation}
H_{\mu\nu}=A_{\nu}\partial_{\mu}(B^{-1/2})-A_{\mu}\partial_{\nu}(B^{-1/2})
\label{e22}
\end{equation}

\noindent which becomes nonzero in regions where the scale factor $B$
changes with position or time. (In the interior and exterior regions of the
bubble $B$ is taken to be approximately constant, but $B$ varies rapidly
with position within the bubble wall.) The effective 4d gauge field
Lagrangian $\mathcal{L}_{4,G}=B^{-1}\mathcal{L}_{G}$ is then given by 
\begin{equation}
\mathcal{L}_{4,G}=-\frac14F^{\mu\nu}F_{\mu\nu}+\frac12B^{-1}\mu_{0}^{2}A^{%
\mu }A_{\mu}-\frac12B^{1/2}F^{\mu\nu}H_{\mu\nu}-\frac14BH^{\mu\nu}H_{\mu\nu }
\label{e23}
\end{equation}

\noindent The gauge boson mass in the effective 4d theory is therefore
identified as 
\begin{equation}
m_{G}=B^{-1/2}\mu_{0}  \label{e24}
\end{equation}

\subsection{Zero Mode Masses}

For the cases of scalar bosons, spin 1/2 Dirac fermions, and massive vector
bosons, it is found that the particle mass in the effective 4d theory is of
the form $m=B^{-1/2}\mu_{0}$, where $\mu_{0}$ is the mass parameter
appearing in the Lagrangian $\mathcal{L}$ of the original 5d theory. Since
the scale factor is assumed to be much larger inside a type I dimension
bubble than outside, i.e., $B_{in}\gg B_{out}$, we have that $m_{in}\ll
m_{out}$ for the $x^{5}$-independent Kaluza-Klein zero modes. Particles
experience a strong, short ranged force within the (thin) wall of a type I
bubble toward the interior. Therefore, particles having nonzero mass tend to
get trapped inside the type I bubble. As the bubble adjusts its size during
equilibration the outward particle pressure on the bubble wall has a
tendency to help stabilize the bubble against total collapse. Just the
opposite holds for a type II bubble, for which $m_{in}\gg m_{out}$.
Particles with nonzero mass are expelled from these bubbles.

\section{Effects of Kaluza-Klein Excitations}

\subsection{Effective 4d Einstein Frame Lagrangian}

Let us now consider the Kaluza-Klein (KK) excitation ($n\neq0$) modes of
scalar bosons, fermions, and massive vector bosons. Each particle mode
contributes a piece to the original 5d Lagrangian of the form 
\begin{equation}
S_{5}=\int d^{5}x\sqrt{\tilde{g}_{5}}\mathcal{L}_{5}=\int d^{4}x\sqrt {-g}%
B^{-1}\int_{0}^{2\pi R}dy\mathcal{L}_{5}  \label{e25}
\end{equation}

\noindent where now $\mathcal{L}_{5}=\mathcal{L}_{5}(x,y)$ and the cylinder
condition is imposed upon the periodic field $\Phi(x,y)=\Phi(x,y+2\pi R)$
allowing the mode expansion 
\begin{equation}
\Phi(x,y)=\sum_{n=-\infty}^{\infty}\Phi_{n}(x)\,e^{iny/R}  \label{e26}
\end{equation}

\noindent Defining $\mathcal{L}=(2\pi R)\mathcal{L}_{5}$ as before, we can
integrate out the $y$ dependence and define 
\begin{equation}
\langle\mathcal{L}\rangle=\frac1{2\pi R}\int_{0}^{2\pi R}dy\,\mathcal{L}%
=\int_{0}^{2\pi R}dy\,\mathcal{L}_{5}  \label{e27}
\end{equation}

\noindent The effective 4d Einstein Frame action then emerges as 
\begin{equation}
S=\int d^{4}x\sqrt{-g}B^{-1}\langle\mathcal{L}\rangle=\int d^{4}x\sqrt {-g}%
\mathcal{L}_{4}  \label{e28}
\end{equation}

\noindent where, as with the case of zero modes, we define the effective 4d
Einstein Frame Lagrangian 
\begin{equation}
\mathcal{L}_{4}=B^{-1}\langle\mathcal{L}\rangle  \label{e29}
\end{equation}

From an expression for the effective 4d Lagrangian for a field we can
identify the effective 4d masses of the KK excitation modes. We again take a
zero mode metric $\tilde{g}_{MN}=\tilde{g}_{MN}(x)$ and examine the scale
factor $B$ dependence of the masses of KK excitations of scalars, spinors,
and vectors.

\subsection{Scalar Bosons}

Consider the scalar boson contribution to the Lagrangian $\mathcal{L}$ given
by 
\begin{equation}
\mathcal{L}_{S}=\tilde{\partial}^{M}\phi^{*}\tilde{\partial}_{M}\phi-\mu
_{0}^{2}\phi^{*}\phi  \label{e30}
\end{equation}

\noindent with 
\begin{equation}
\phi(x,y)=\sum_{n}\phi_{n}(x)\,e^{iny/R}  \label{e31}
\end{equation}

\noindent The Lagrangian $\mathcal{L}_{S}$ can then be written in terms of
the KK modes as 
\begin{equation}
\mathcal{L}_{S}=\sum_{m,n}\left\{ (\tilde{\partial}^{\mu}\phi_{m}^{*})(%
\tilde{\partial}_{\mu}\phi_{n})-\left( \frac{mn}{B^{2}R^{2}}%
+\mu_{0}^{2}\right) \phi_{m}^{*}\phi_{n}\right\} e^{i(n-m)y/R}  \label{e33}
\end{equation}

\noindent Using 
\begin{equation}
\frac1{2\pi R}\int_{0}^{2\pi R}dy\,e^{i(n-m)y/R}=\delta_{mn}  \label{e34}
\end{equation}

\noindent we obtain 
\begin{equation}
\langle\mathcal{L}_{S}\rangle=\sum_{n}\left\{
B(\partial^{\mu}\phi_{n}^{*})(\partial_{\mu}\phi_{n})-\left( \frac{n^{2}}{%
B^{2}R^{2}}+\mu_{0}^{2}\right) |\phi_{n}|^{2}\right\}  \label{e35}
\end{equation}

\noindent The effective 4d scalar Lagrangian $\mathcal{L}_{4,S}=B^{-1}\langle%
\mathcal{L}_{S}\rangle$ is then 
\begin{equation}
\mathcal{L}_{4,S}=\sum_{n}\left\{ (\partial^{\mu}\phi_{n}^{*})(\partial_{\mu
}\phi_{n})-\left( \frac{n^{2}}{B^{3}R^{2}}+\frac{\mu_{0}^{2}}B\right)
|\phi_{n}|^{2}\right\}  \label{e36}
\end{equation}

\noindent and the mass of the $n^{th}$ KK scalar boson excitation in the
effective 4d theory is 
\begin{equation}
m_{S,n}=\left( \frac{\mu_{0}^{2}}B+\frac{n^{2}}{B^{3}R^{2}}\right) ^{1/2}
\label{e37}
\end{equation}

\subsection{Fermions}

Consider the fermionic Lagrangian 
\begin{equation}
\mathcal{L}_{F}=\bar{\Psi}^{\prime}\left( i\Gamma^{M}\partial_{M}-\mu
_{0}\right) \Psi  \label{e38}
\end{equation}

\noindent where $\Psi=\Psi(x,y)$, $\bar{\Psi}^{\prime}=\Psi^{\dagger}%
\Gamma^{0}$, and the $\Gamma$ matrices are normalized according to $\left\{
\Gamma^{M},\Gamma^{N}\right\} =-2\tilde{g}^{MN}$, so that 
\begin{equation}
\left\{ \Gamma^{\mu},\Gamma^{\nu}\right\} =-2\tilde{g}^{\mu\nu}=-2Bg^{\mu
\nu},\,\,\,\,\,\,\,(\Gamma^{55})^{2}=-(\tilde{g}^{55})^{2}=B^{-2}
\label{e39}
\end{equation}

\noindent As before, in order to pass to the effective 4d theory we define a
set of $\gamma$ matrices by 
\begin{equation}
\Gamma^{\mu}=B^{1/2}\gamma^{\mu},\,\,\,\,\,\,\,\,\,\,\Gamma^{5}=B^{-1}%
\gamma^{5},\,\,\,\,\,\gamma^{5}=\left( 
\begin{array}{cc}
-1 & 0 \\ 
0 & 1%
\end{array}
\right)  \label{e40}
\end{equation}

\noindent satisfying $\left\{ \gamma^{\mu},\gamma^{\nu}\right\} =-2g^{\mu
\nu}$ and $(\gamma^{5})^{2}=1$. In terms of the $\gamma$ matrices, the
Lagrangian becomes 
\begin{equation}
\begin{array}{cc}
\mathcal{L}_{F} & =B^{1/2}\bar{\Psi}\left( B^{1/2}i\gamma^{\mu}\partial_{\mu
}+B^{-1}i\gamma^{5}\partial_{5}-\mu_{0}\right) \Psi \\ 
& =B\bar{\Psi}\left(
i\gamma^{\mu}\partial_{\mu}+B^{-3/2}i\gamma^{5}\partial_{5}-B^{-1/2}\mu_{0}%
\right) \Psi%
\end{array}
\label{e41}
\end{equation}

\noindent where $\bar{\Psi}=\Psi^{\dagger}\gamma^{0}=B^{-1/2}\bar{\Psi }%
^{\prime}$. In the dimensionally reduced 4d theory the term proportional to $%
\bar{\Psi}(i\gamma^{5}\partial_{5})\Psi$ corresponds to a mass term, so that
we require $\Psi$ to be an eigenfunction of $i\gamma^{5}\partial_{5}$.

The field $\Psi$ must therefore satisfy the periodicity condition $%
\Psi(x,y)=\Psi(x,y+2\pi R)$ and be an eigenfunction of $i\gamma^{5}%
\partial_{5}$. Let us introduce a chiral notation and write the field $\Psi$
in the form 
\begin{equation}
\Psi(x,y)=\sum_{n=-\infty}^{\infty}\Psi_{n}(x,y)=\sum_{n=-\infty}^{\infty
}\left( 
\begin{array}{c}
\psi_{nL}(x)\xi_{nL}(y) \\ 
\psi_{nR}(x)\xi_{nR}(y)%
\end{array}
\right) ,\,\,\,\,\,\left( 
\begin{array}{cc}
\xi_{nL}= & e^{i\alpha_{L}|n|y/R} \\ 
\xi_{nR}= & e^{i\alpha_{R}|n|y/R}%
\end{array}
\right)  \label{e42}
\end{equation}

\noindent where $\alpha_{L,R}$ each take a value of $\pm1$ in order to
satisfy both the periodicity and eigenvalue conditions. Using the form of $%
\gamma^{5}$ given by eq. (\ref{e40}), the eigenvalue condition 
\begin{equation}
i\gamma^{5}\partial_{5}\Psi_{n}=\lambda_{n}\Psi_{n}  \label{e43}
\end{equation}

\noindent yields $\alpha_{L}=-1$, $\alpha_{R}=+1$, and $\lambda_{n}=-|n|/R$.
We therefore have 
\begin{equation}
\Psi_{n}=\left( 
\begin{array}{c}
\psi_{nL}(x)\,e^{-i|n|y/R} \\ 
\psi_{nR}(x)\,e^{i|n|y/R}%
\end{array}
\right)  \label{e44}
\end{equation}

We can perform the integration of $\mathcal{L}_{F}$ over $y$ to obtain $%
\langle\mathcal{L}_{F}\rangle$. The orthogonality of the $\xi_{n}(y)$
functions can be used, and for a specific representation of $\gamma$
matrices let us use, e.g., $-i\gamma^{\mu}=\left( 
\begin{array}{cc}
0 & \sigma^{\mu} \\ 
\bar{\sigma}^{\mu} & 0%
\end{array}
\right) $. Then for the effective 4d Lagrangian $\mathcal{L}%
_{4,F}=B^{-1}\langle\mathcal{L}_{F}\rangle$, we get 
\begin{equation}
\mathcal{L}_{4,F}=\sum_{n}\bar{\psi}_{n}\left( i\gamma^{\mu}\partial_{\mu
}-m_{F,n}\right) \psi_{n}  \label{e45}
\end{equation}

\noindent where 
\begin{equation}
\psi_{n}(x)=\left( 
\begin{array}{c}
\psi_{nL}(x) \\ 
\psi_{nR}(x)%
\end{array}
\right)  \label{e45a}
\end{equation}

\noindent The mass of the $n^{th}$ KK excitation in the effective 4d theory
is 
\begin{equation}
m_{F,n}=\frac{\mu_{0}}{B^{1/2}}+\frac{|n|}{B^{3/2}R}  \label{e46}
\end{equation}

\subsection{Massive Vector Bosons}

As in the zero mode description, the contribution to the Lagrangian from a
U(1) gauge field $A_{M}^{\prime}$ that acquires a mass described by the
parameter $\mu_{0}$ in the 5d theory is 
\begin{equation}
\mathcal{L}_{G}=-\frac14\tilde{F}^{\prime MN}\tilde{F}_{MN}^{\prime}+\frac12%
\tilde{g}^{MN}\mu_{0}^{2}A_{M}^{\prime}A_{N}^{\prime}  \label{e47}
\end{equation}

\noindent where now $A_{M}^{\prime}=A_{M}^{\prime}(x,y)$. As before, we set $%
A_{5}^{\prime}=0$ and introduce the field $A_{\mu}=B^{1/2}A_{\mu}^{\prime}$,
with the periodic field $A_{\mu}$ satisfying the periodicity requirement
with a mode expansion 
\begin{equation}
A_{\mu}(x,y)=\sum_{n=-\infty}^{\infty}A_{\mu}^{n}(x)\,e^{iny/R}  \label{e48}
\end{equation}

\noindent Upon rewriting the Lagrangian in terms of the 4d metric $%
g_{\mu\nu} $ and the field $A_{\mu}$, $\mathcal{L}_{G}$ assumes the form 
\begin{equation}
\begin{array}{ll}
\mathcal{L}_{G}= & -\frac14BF^{\mu\nu}F_{\mu\nu}+B^{-2}F_{\mu5}F_{\,\,\,%
\,5}^{\mu}-\frac12B^{3/2}H_{\mu\nu}F^{\mu\nu} \\ 
& -\frac14B^{2}H_{\mu\nu}H^{\mu\nu}+\frac12\mu_{0}^{2}A_{\mu}A^{\mu}%
\end{array}
\label{e49}
\end{equation}

\noindent where $H_{\mu\nu}$ is defined in eq.(\ref{e22}). Inserting the
mode expansion and integrating out the $y$ dependence leaves 
\begin{equation}
\begin{array}{ll}
\langle\mathcal{L}_{G}^{(n)}\rangle= & -\frac14BF_{\mu\nu}^{*n}F^{n\mu\nu
}+\frac12\left( \frac{n^{2}}{B^{2}R^{2}}+\mu_{0}^{2}\right)
A_{\mu}^{*n}A^{n\mu} \\ 
& -\frac12B^{3/2}H_{\mu\nu}^{*n}F^{n\mu\nu}-\frac14B^{2}H_{\mu\nu}^{*n}H^{n%
\mu\nu}%
\end{array}
\label{e50}
\end{equation}

\noindent where $F_{\mu\nu}^{n}=\partial_{\mu}A_{\nu}^{n}-\partial_{\nu}A_{%
\mu}^{n} $, $A_{\mu}^{*n}=A_{\mu}^{-n}$, etc., and $\langle\mathcal{L}%
_{G}\rangle=\sum_{n}\langle\mathcal{L}_{G}^{(n)}\rangle$. The effective 4d
Lagrangian $\mathcal{L}_{4,G}=B^{-1}\langle\mathcal{L}_{G}\rangle$ is
therefore given by $\mathcal{L}_{4,G}=\sum_{n}\langle\mathcal{L}%
_{4,G}^{(n)}\rangle$, with 
\begin{equation}
\begin{array}{ll}
\mathcal{L}_{4,G}^{(n)}=B^{-1}\langle\mathcal{L}_{G}^{(n)}\rangle= & 
-\frac14F_{\mu\nu}^{*n}F^{n\mu\nu}+\frac12\left( \frac{n^{2}}{B^{3}R^{2}}+%
\frac{\mu_{0}^{2}}B\right) A_{\mu}^{*n}A^{n\mu\nu} \\ 
& -\frac12B^{1/2}H_{\mu\nu}^{*n}F^{n\mu\nu}-\frac14BH_{\mu\nu}^{*n}H^{n\mu%
\nu}%
\end{array}
\label{e51}
\end{equation}

\noindent From this we identify the mass of the $n^{th}$ KK vector boson
mode appearing in the effective 4d theory as 
\begin{equation}
m_{G,n}=\left( \frac{\mu_{0}^{2}}B+\frac{n^{2}}{B^{3}R^{2}}\right) ^{1/2}
\label{e52}
\end{equation}

\subsection{Kaluza-Klein Excitation Masses}

We see from the above that the mass of the $n^{th}$ KK excitation has the
form 
\begin{equation}
m_{n}=\left\{ 
\begin{array}{ll}
\left( \dfrac{\mu_{0}^{2}}B +\dfrac{n^{2}}{B^{3}R^{2}} \right) ^{1/2}, & 
\text{bosons} \\ 
\dfrac{\mu_{0}}{B^{1/2}} +\dfrac{|n|}{B^{3/2}R} , & \text{fermions}%
\end{array}
\right\}  \label{e52a}
\end{equation}

\noindent where $\mu_{0}$ is the mass parameter appearing in the 5d theory,
and $m_{0}=\mu_{0}/B^{1/2}$ is the zero mode mass. If the zero mode mass
vanishes, then $m_{n}=\frac{|n|}{B^{3/2}R}$, and, in this case, $\frac {%
m_{n,out}}{m_{n,in}}=\left( \frac{B_{in}}{B_{out}}\right) ^{3/2}$. For a
type I bubble where $B_{in}/B_{out}\gg1$, KK modes which may be too massive
to be produced outside the bubble may be produced in the bubble's interior
and can therefore help to stabilize the bubble against collapse. However,
for a type II bubble where $B_{in}/B_{out}\ll1$, the KK modes would be
expelled from the bubble's interior.

\section{Radiation Stabilized Bubbles}

Let us consider a type I bubble with a high temperature interior that
contains radiation modes comprised of photons as well as particles with
masses $m_{in}\ll|\vec{p}|$. For a type II bubble, we take the limit where
there are only photons inside. (In a type I bubble, there may be
nonrelativistic heavy KK states as well, for example, but the energy density
is assumed to be negligible in comparison to that of the relativistic
species.) The mass of a particle outside of the type I bubble is $m_{out}\gg
m_{in}$ for $B_{out}\ll B_{in}$, but particles in the bubble interior having
energy $\omega\geq m_{out}$ can escape the bubble. The relative number of
particles that escape over the lifetime of the bubble depends upon $%
m_{out}/T=\beta m_{out}$, assuming a thermal distribution of energies. (For $%
\beta m_{out}\gg1$ the fractional number of particles escaping the bubble
will be small.) Let us assume that there is an effective number of spin
degrees of freedom $\bar
{g}=\bar{g}_{B}+\frac78\bar{g}_{F}$ associated
with radiation modes inside the bubble that tend to stabilize the bubble
without quickly escaping, i.e., $\beta m_{out}\gg1$. The photons inside the
bubble escape at a very low rate as well\cite{GM}, due to the high photon
opacity of the bubble wall. [The transmission coefficient for photons\cite%
{GM} is roughly $\mathcal{T}\sim O(B_{<}/B_{>})\ll1$, where $B_{<}$ ($B_{>}$%
) is the smaller (larger) of $B_{in}$ or $B_{out}$.] We therefore expect a
bubble with a variety of radiation modes to be a metastable state with a
lifetime $\tau$ that is at least as long as that of a photon stabilized
bubble\cite{GM}, i.e., $\tau\gtrsim R_{B,0}/\mathcal{T}$, where $R_{B,0}$ is
the initial radius of the bubble.

We can estimate the equilibrium radius and mass of a stabilized bubble with
an interior temperature $T$. We take the radiation energy density inside to
be\footnote{%
It is assumed here, for simplicity, that not too many Kaluza-Klein modes are
excited, so that we can use the familiar 4d expressions for radiation energy
densities, etc. This condition is satisfied if, for instance, $T<(BR)^{-1}$,
where $R$ is the radius parameter associated with the extra dimension. (See,
e.g., \cite{RSS,Kolb}.)} 
\begin{equation}
\rho _{Rad}=AT^{4}=\bar{g}\frac{\pi ^{2}}{30}T^{4}  \label{e53}
\end{equation}

\noindent In addition, there may be a volume term $\mathcal{E}%
_{V}=\frac43\pi R_{B}^{3}\lambda$, where the value of the effective
potential in the bubble's interior, $U(\varphi)=\lambda$, is taken to be a
constant. Using this approximation, along with a thin wall approximation for
the bubble wall, the expression for the bubble mass becomes 
\begin{equation}
M=\mathcal{E}_{Rad}+\mathcal{E}_{V}+\mathcal{E}_{Wall}=\frac43\pi
R_{B}^{3}(\rho_{Rad}+\lambda)+4\pi R_{B}^{2}\sigma  \label{e54}
\end{equation}

\noindent where $\sigma$ is the surface energy density of the wall. We
further assume that the bubble, after it forms, will adjust its radius to
reach an equilibrium state with an approximately adiabatic (isentropic)
expansion or contraction. The radiation entropy density is $s\sim T^{3}$, so
that if the bubble reaches equilibrium on a relatively small time scale, we
can take $R_{B}T=const$ during the equilibration process. With these
assumptions we obtain approximate expressions for an equilibrium radius $%
R_{B}$ and an equilibrium mass $M$ given by 
\begin{equation}
R_{B}=\frac{6\sigma}{(AT^{4}-3\lambda)}=\frac{6\sigma}{(\rho_{Rad}-3\lambda )%
}\,  \label{e55a}
\end{equation}

\noindent and 
\begin{equation}
M=2\pi R_{B}^{3}(\rho_{Rad}-\tfrac13 \lambda)=12\pi\sigma R_{B}^{2}\left( 
\frac{\rho_{Rad}-\frac13\lambda}{\rho_{Rad}-3\lambda}\right)  \label{e55b}
\end{equation}

\noindent For a bubble that rapidly adjusts its size to reach equilibrium in
an adiabatic manner so that, approximately, $R_{B}T=const$ during this
adjustment period, a collapsing bubble rapidly heats up and an expanding
bubble rapidly cools down, but the final temperature of the stabilized
bubble must satisfy $\rho_{Rad}=AT^{4}>3\lambda$. On the other hand, it is
assumed that the bubble's domain wall forms at some temperature $T_{c}$
(which will depend upon model parameters) where a barrier forms in the
effective potential to separate the two low energy states, so that a bubble
at equilibrium should have an interior temperature $T<T_{c}$. In summary,
for a stabilized bubble with radius and mass given by (\ref{e55a}) and (\ref%
{e55b}), respectively, the interior temperature must lie within the range 
\begin{equation}
\left( \frac{3\lambda}A\right) ^{1/4}<T<T_{c}  \label{e55c}
\end{equation}

The lifetime of the bubble depends upon the rate at which photons and other
high energy particles inside the bubble escape. This, in turn, depends upon
the interactions of the particles inside the bubble and the rates at which
lighter particles with $\omega>m_{out}$ are produced, and the rates at which
they escape. However, we can reason that the lifetime of a bubble will be at
least as long as that of a bubble stabilized by photons alone, in which case%
\cite{GM} $\tau\gtrsim R_{B,0}/\mathcal{T}\sim\mathcal{O}%
(B_{>}/B_{<})R_{B,0} $.

If a bubble continues to shrink and the interior temperature increases above
the critical temperature $T_{c}$ (corresponding to the temperature at which
the domain wall forms), the wall disappears, i.e., the bubble explodes. When
the bubble explodes, its radiative contents consisting of high temperature
photons, bosons, and fermions are suddenly released.

\section{4d Gravitational Aspects}

We now turn attention to consider 4d gravitational aspects of dimension
bubbles and argue that, under rather mild assumptions, the interior metric
of a dimension bubble shows that its geometrical center is singularity free.
This is in contrast to the situation found for a ``gravitational bag''\cite%
{DG}, which is basically an ``empty'', static idealization of a type I
dimension bubble. In particular, the gravitational bag possesses a naked
singularity at its center\cite{DG,Guen}. (The gravitational bag considered
in ref.\cite{DG} arises from a Freund-Rubin compactification of a 6d theory,
but has the same essential aspects that we have in our 5d model. The \textit{%
exact} solution obtained for the gravitational bag metric assumes that the
effective potential vanishes in the bag's interior, corresponding to the
extra dimension becoming infinitely large ($\varphi,B\rightarrow\infty$)
there.) The gravitational bag contains only the scalar field $\varphi$ that
gives rise to the bubble wall, without any entrapped particles in the
bubble's interior. We argue that, unlike a gravitational bag, the fluidic
interior of a dimension bubble does not exhibit a naked singularity at its
center.

The interior 4d geometry of the gravitational bag is described by\cite%
{DG,Guen} 
\begin{equation}
ds^{2}=C_{1}\left( \frac{ar-1}{ar+1}\right) ^{2p}dt^{2}-\frac{C_{2}}{%
a^{4}r^{4}}\frac{(ar+1)^{2(p+1)}}{(ar-1)^{2(p-1)}}(dr^{2}+r^{2}d\Omega ^{2})
\label{e56}
\end{equation}

\noindent where $r$ is an isotropic radial coordinate and $C_{1}$, $C_{2}$, $%
a$, and $p$ are constants, with\cite{DG} $\frac12<p<1$. The center of
geometry is located at $r=1/a$, where the coefficient of $d\Omega^{2}$ goes
to zero as $r\rightarrow1/a$ from above, and the metric spawns a singularity
there, corresponding to a naked singularity of the exact solution\cite{DG}.
(See also ref.\cite{Guen}.) In ref.\cite{DG} it is found that although there
is a singularity at the bag's center, the mass of the gravitational bag is
finite.

Now, instead of a gravitational bag, consider a dimension bubble that is
filled with particles trapped in the interior. We assume that the
stress-energy of the $\varphi$ field in the interior is negligible in
comparison to that associated with the particles, and, furthermore, that we
can approximate the bubble's contents by an isothermal fluid with pressure $%
p $ and energy density $\rho$ connected by $p=w\rho$, with $w$ a constant.
(For the case of radiation, $w=1/3$.) In this case, the 4d fluid-filled
domain wall bubble can be viewed as a \textit{cosmic balloon}\cite{BH,wang}.
The geometry inside the cosmic balloon is described by 
\begin{equation}
ds^{2}=B(r)dt^{2}-A(r)dr^{2}-r^{2}d\Omega^{2}  \label{e57}
\end{equation}

\noindent where the coordinate $r$ is now a nonisotropic radial coordinate
(the metric coefficient $B(r)$ used here is not to be confused with $|\tilde{%
g}_{55}|^{1/2}$). The fluid in the bubble's interior has the important
effect of removing the singularity at the bubble's center, $r=0$, provided
that the fluid density $\rho(r)$ remains finite at the bubble's center. This
can be seen by borrowing some of the results obtained by Wang (see \cite%
{wang} and references therein.) Specifically, near the center we have 
\begin{equation}
A(r)=\left[ 1-x(r)\right] ^{-1}\approx\left[ 1-\frac83\pi G\rho (0)r^{2}%
\right] ^{-1}  \label{e58}
\end{equation}

\noindent where $\rho(0)$ is the central density and $x(r)\approx\frac83\pi
G\rho(0)r^{2}$. For $\rho(0)$ finite, we therefore have $A(0)=1$, and a
naked singularity at the center is avoided. The coefficient $B(r)$ can be
obtained from the equilibrium condition 
\begin{equation}
\frac{B^{\prime}}B=-\frac{2p^{\prime}}{p+\rho}=-\left( \frac{2w}{1+w}\right) 
\frac{\rho^{\prime}}\rho  \label{e59}
\end{equation}

\noindent where $^{\prime}$ denotes differentiation with respect to $r$.
Defining $\bar{t}=\left[ 8\pi G\rho(0)\right] ^{1/2}r$, we have $\frac1B%
\frac{dB}{d\bar{t}}\approx\left( \frac{4w}{1+w}\right) \bar{t}$ for $\bar{t}%
\ll1$. Integration gives 
\begin{equation}
B(r)\approx B(0)\exp\left\{ \left( \frac{2w}{1+w}\right) \left( 8\pi
G\rho(0)r^{2}\right) \right\}  \label{e60}
\end{equation}

Fluid-filled dimension bubbles, including radiation stabilized bubbles,
therefore seem to have interior gravitational fields that are better behaved
near the centers than those of gravitational bags.

\section{Summary}

An inhomogeneous compactification of a higher dimensional spacetime to 4d
may result in the formation of ``dimension bubbles''. The sizes of the extra
dimensions inside the bubble can be either larger (for type I) or smaller
(for type II) than outside the bubble. Whether a dimension bubble is type I
or type II depends upon the form of the effective potential $U(\varphi)$
controlling the size of the extra dimension. We can therefore have the
situation where a type I dimension bubble encloses a ``large'' extra
dimension and is surrounded by a ``small'' extra dimension, with the
opposite holding true for a type II bubble. For example, if the size of an
extra dimension $BR$ varies between a Planck size and a TeV size, the extra
dimensional scale factor $B$ can change by 16 orders of magnitude. As a
result of a dramatic variation of the size of the extra dimension across the
bubble wall, the masses of various particle modes can vary dramatically
between the inside and outside of the bubble, leading to an enhanced
stabilization of a type I bubble by the entrapment of lighter particles
inside or the expulsion of heavier particles from inside a type II bubble.
The dependence of the particle mass upon the extra dimensional scale factor $%
B$ has been demonstrated for bosons and fermions, including both
Kaluza-Klein (KK) zero modes and KK excitation modes. [See eq.(\ref{e52a}).]

Both types of bubbles can be stabilized by photons, due to the high
reflectivity of the bubble wall, and either type of dimension bubble may
exist as a long-lived metastable state. Some basic features of plasma-filled
bubbles have been examined and estimates obtained for the equilibrium radius
and mass. The lifetime of a bubble with an initial radius $R_{B,0}$ is
roughly estimated to be $\tau\gtrsim\mathcal{O}(B_{>}/B_{<})R_{B,0}$, where $%
B_{>}$($B_{<}$) is the larger (smaller) value of $B$ on the inside or
outside of the bubble.

Finally, we have considered the 4d gravitational aspects of dimension
bubbles to argue that, unlike the case for a \textquotedblleft gravitational
bag\textquotedblright \cite{DG,Guen} (which may be thought of as an empty,
static idealization of a type I dimension bubble) the center of a
fluid-filled dimension bubble is singularity free. This is seen by treating
the dimension bubble as a \textit{cosmic balloon}\cite{BH} and borrowing the
results for the interior metric of a cosmic balloon\cite{wang}. Fluid-filled
dimension bubbles are therefore seen to have better behaved interior
gravitational fields than those of gravitational bags.

Since the existence of dimension bubbles depends upon a dramatic \textit{%
change} in the sizes of extra dimensions across the bubble wall rather than
upon the actual size of an extra dimension in any region, the detections of
such solitons could provide evidence for the existence of extra dimensions,
regardless of their sizes.

\smallskip

\smallskip\ 

\ \textbf{Acknowledgement}

I wish to thank E.I. Guendelman for valuable comments.

\smallskip\ 

\smallskip\ 

\smallskip\ 

\smallskip\

\end{document}